# ZnPSe$_3$ as ultrabright indirect bandgap system with microsecond excitonic lifetimes.

M. Grzeszczyk[1,2], K. S. Novoselov[1,2], M. Koperski[1,2,*]

[1]Department of Materials Science and Engineering, National University of Singapore, 117575, Singapore

[2]Institute for Functional Intelligent Materials, National University of Singapore, 117544, Singapore

E-mail: * msemaci@nus.edu.sg

**We report an optical characterization of ZnPSe$_3$ crystals that demonstrates indirect band gap characteristics in combination with unusually strong photoluminescence. We found evidence of interband recombination from excitonic states with microsecond lifetimes. Through optical characterization, we reconstructed the electronic band scheme relevant for fundamental processes of light absorption, carrier relaxation and radiative recombination. The investigation of the radiative processes in the presence of magnetic field revealed spin polarization of fundamental electronic states. This observation opens a pathway towards controlling the spin of excitonic states in technologically relevant microsecond timescales.**

The band structure theory of semiconductors introduces the concept of a valley, which exists as an extremum in the valence or conduction band. Most commonly, multiple valleys are found in band structure of semiconductors and insulators. They can host electrons and holes partaking in processes that define structural, optical, electrical or magnetic properties of crystals. Charge carriers residing at various valleys can hold very different properties, as they are characterized by their specific effective mass and orbital/spin composition of wave functions[1]. Creating, detecting and manipulating excitations within particular valleys is desirable for gaining fundamental understanding of the unique characteristics of charge carriers in various types of materials as well as enabling their novel functionalities. For that reason, a discipline of valleytronics[2] was established, treating the valley degree of freedom as a quantum number that determines the behavior of charge carriers.

Early exploration of valleytronics was conducted for silicon and germanium[3], where the accessibility of valleys is limited by their weak optical response. Recently, non-trivial valley structure was uncovered in representatives of two dimensional materials[4,5]. Particularly interesting for opto-electronic applications[6,7] are monolayer transition metal dichalcogenides (TMDC)[8-10], which were found to display robust luminescence as direct band gap systems in monolayer limit and significantly suppressed luminescence as indirect band gap systems in multilayer limit[11-16]. That is a common finding that indirect excitonic recombination is often quenched by non-radiative processes. The long lifetime of indirect states puts the radiative recombination in competition with non-radiative processes, which usually occur within shorter timescales.

Here, we demonstrate that ZnPSe$_3$ acts as two-dimensional indirect band gap semiconductor while displaying robust yellow photoluminescence (PL). The comparative study of PL and quasi-absorption characteristics unveils optical resonances due to selective excitations at $\Gamma$ and K points of the Brillouin zone and radiative recombination from indirect $\Gamma$-K transition. The lifetime of the fundamental excitonic state is found to be in microsecond regime, about six orders of magnitude longer than in monolayer TMDC[17,18]. In magnetic field, a polarized emission is observed for the radiative transition indicative of spin splitting which is preserved throughout the lifetime of the exciton.

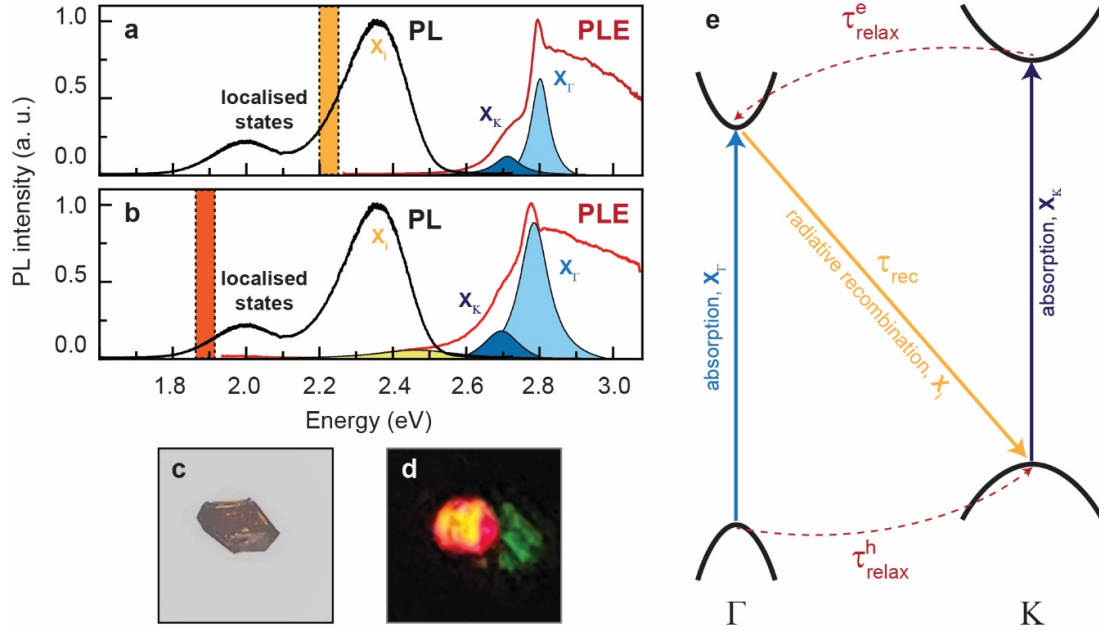

**Figure 1. PL and PLE characterisation of ZnPSe$_3$ crystals**. The low temperature (1.6 K) PL spectra under 3.06 eV laser excitation were measured in back-scattering microscopic geometry. They are confronted with PLE spectra for higher energy **(a)** and lower energy **(b)** PL bands. For PLE spectra, the intensity of the PL is integrated at the lower energy tail for both bands, as highlighted by stripes in parts **(a)** and **(b)**. The yellow/orange PL is observable with a naked eye in ambient conditions as illustrated by photographs of the crystal under white light illumination **(c)** and under 3.06 eV macroscopic laser excitation in darkness **(d)** with a longpass filter placed in front of the camera to remove stray laser light. The comparative analysis of the PL and PLE data allow us to construct a minimal diagram of light absorption, carrier relaxation and radiative recombination processes **(e)** occurring within ZnPSe$_3$ crystals.

We firstly inspected the optical response of the ZnPSe$_3$ crystal at the temperature of 1.6 K. We found that the PL spectrum is composed of two distinct emission bands at 2.00 eV and 2.36 eV as presented in **Fig. 1(a,b)**. Both bands display similar photoluminescence excitation (PLE) spectra which reveal multiple contributions to the processes of light absorption. We performed the deconvolution of the excitation spectrum into 3 components: 1) direct interband absorption edge characterized by three dimensional density of states and excitonic absorption resonances labeled 2) $X_K$ and 3) $X_\Gamma$ in **Fig 1(a,b)**. The analysis of the PLE spectrum enables estimation of the exciton binding energy for $X_\Gamma$ to be 19 ± 8 meV and for $X_K$ to be 110 ± 10 meV based on a fitting procedure described in **S. I. Appendix**. The comparative analysis of the PL and PLE spectra establishes ZnPSe$_3$ as an indirect bandgap system with a separation between the highest energy emission band and the absorption edge of about 0.4 eV. In contrast with a common observation that indirect bandgap materials are characterized by weak photoluminescence, ZnPSe$_3$ crystals are unusually bright. The yellow photoluminescence is observable with a naked eye in ambient conditions as demonstrated in **Fig. 1(c,d)**.

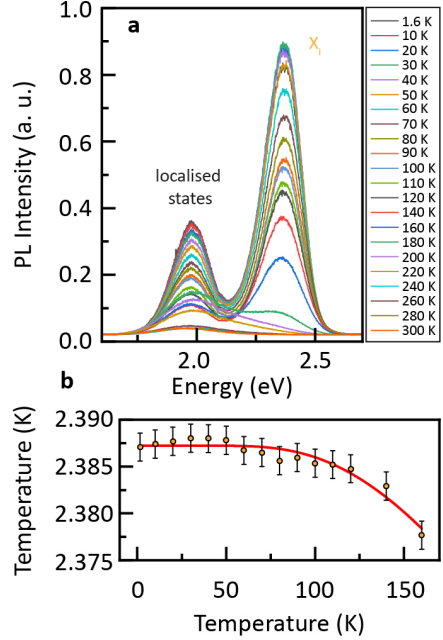

**Figure 2.** The temperature dependence of the PL spectra is inspected for resonant $\Gamma$ excitation in the range of temperatures from 1.6 K to 300 K **(a)**. The redshift of the emission energy of the higher energy band **(b)** is presented based on Gaussian fitting of the emission resonance in the temperature range where the higher energy band remains a distinctive feature of the PL spectrum.

The comparative analysis of PL and PLE spectra allows us to identify the processes of light absorption, photo-exited carrier relaxation and radiative recombination. Based on the density functional theory (DFT) band structure[19-21], the fundamental band gap occurs between the minimum of conduction band at $\Gamma$ point and maximum of the valence band at the K point of the Brillouin zone. In a high quality crystal, the interband $\Gamma$-K recombination should constitute a dominant radiative channel and create optical resonances at highest energy. For those reasons we attribute the PL band at 2.36 eV to the indirect exciton recombination. The DFT further predicts that there are two direct gaps at $\Gamma$ and K point separated by small energy difference (within 40 meV). Indeed, we observe two excitonic resonances in the PLE spectra at 2.71 eV and 2.80 eV. Notably, the resonances differ significantly by their linewidths of 91 ± 4 meV and 64 ± 3 meV. The linewidth of an absorption resonance is directly linked with the lifetime of the exciton by the Heisenberg uncertainty principle. Typically, electronic states exhibiting spatial localization display longer lifetimes leading to the broadening of the optical resonances. In this view, the K-valley charge carriers, which are built predominantly from the d-orbitals of the Zn atom, are localized in the real space and the interband optical transition is characterized by large energy uncertainty leading to broad linewidth. Similarly, $\Gamma$-valley charge carriers are composed of s- and p-type wave functions of P and Se atoms, which are delocalized in real space hence raising narrow absorption resonances. Therefore, we identify the two absorption resonances as $X_\Gamma$ and $X_K$ which correspond to exitons at $\Gamma$ and K point respectively.

From the analysis of the optical resonances in emission and quasi-absorption spectra, we established a diagram of absorption, carrier relaxation and radiative recombination processes that determine the optical response of ZnPSe$_3$ crystals (see **Fig 1(e)**). The absorption processes create excitons at $\Gamma$ and K point of the Brillouin zone followed by a relaxation of an electron for the K excitation or a hole for $\Gamma$ excitation onto the fundamental indirect state. The relaxation time of the electron and the hole may be seen as a lifetime of direct K and Γ excitons, respectively. They can be estimated from the linewidth of the absorption resonances ($\tau \approx \frac{\hbar}{2\Delta E}$) to be $\tau_{relax}^e \approx 22\ fs$ and $\tau_{relax}^h \approx 32\ fs$.

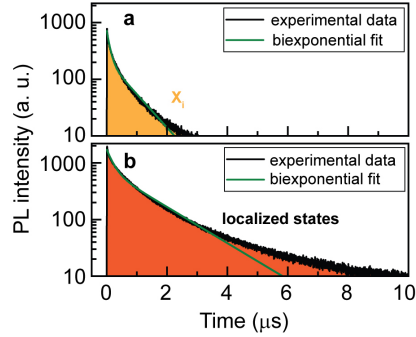

**Figure 3. PL decay transients in ZnPSe₃ crystals.** The PL decay transients have been inspected for higher energy **(a)** and lower energy **(b)** PL bands under 2.41 eV excitation with a picosecond pulsed laser. Both PL transients were fitted with biexponential decay functions: $I(t) = A_1 e^{-t/\tau_1} + A_2 e^{-t/\tau_2}$ with characteristic decay times $\tau_1$ and $\tau_2$.

The lower energy (2.00 eV) band is attributed to recombination of donor-bound excitons based on the inspection of power dependence of the emission intensity presented in S. I. Appendix in **Fig. S1**. The identification of the free exciton band and donor-bound exciton band is further supported by the characterization of the temperature dependence of the PL spectra. Characteristically for semiconductors with radiative ground state, the PL intensity from ZnPSe₃ crystals decreases (**Fig. 2(a)**) with the raise of temperature due to activation of non-radiative channels and thermal depopulation of the ground state. The donor-bound excitons are characterized by larger binding energy hence they are less affected by the thermal effects. Indeed, we observe that the donor-bound exciton band becomes the dominant optical resonance at about 180 K. We also observe a redshift for both emission bands, that we analyze quantitatively for the free exciton band in **Fig 2(b)**. We assume that the major contribution to the temperature induced band gap reduction originates from the coupling of electrons forming the crystal bonds with phonons[22,23]. In such case, the band gap dependence on temperature may be expressed as:

$$E_g(T) = E_g(0) - S\langle\hbar\omega\rangle\left(\coth\left(\frac{\langle\hbar\omega\rangle}{2kT}\right) - 1\right)$$

where $E_g(0)$ denotes the value of the band gap at 0 K, $S$ is the electron-phonon coupling constant, $\langle\hbar\omega\rangle$ is the mean phonon energy and $kT$ is the thermal energy. From the fitting of this formula to the temperature dependence of the free exciton band energy, we find that $S = 0.0024 \pm 0.0011$ and $\langle\hbar\omega\rangle = 44.3 \pm 8.3\ meV$. The value of the mean phonon energy corresponds well to the optical phonon resonances[24] $E_g$ and $A_{1g}$ in ZnPSe₃ at 52.9 meV and 57.4 meV demonstrating that it is the phonon effects that are mostly responsible for the band gap reduction. Additional band gap modifications could originate from thermal expansion and/or alteration of the exciton binding energy with temperature.

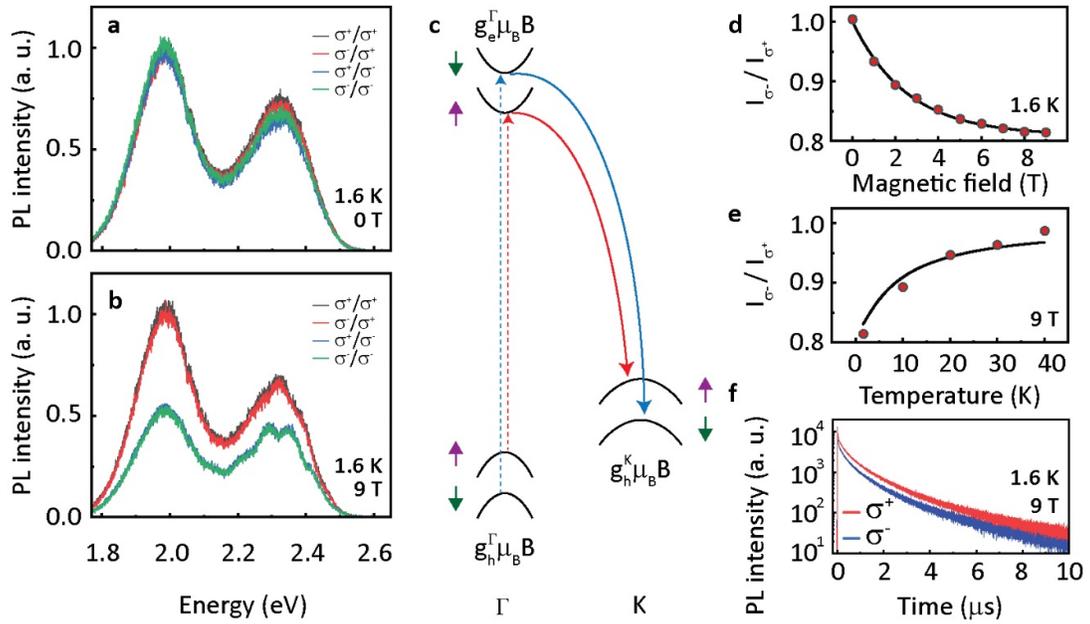

**Figure 4. The magneto-PL characterisation of ZnPSe₃ crystals.** The PL spectra were resolved by the circular polarisation of the excitation and emitted photons (with the notation excitation/detection circular polarisation) under the condition of resonant $\Gamma$ excitation. The PL spectra for four possible polarization configurations are presented for 0 T **(a)** and 9 T **(b)** with the magnetic field applied perpendicularly to the surface of the crystal in Faraday configuration. The observed polarisation properties are interpreted in terms of linear Zeeman splitting (quantified by band-specific g-factors) of fundamental conduction and valence bands at $\Gamma$ and K points of the Brillouin zone as illustrated by the schematic **(c)**. The polarisation degree for the higher energy band is measured as a function of magnetic field **(d)**, temperature **(e)**. The polarisation-dependent PL transients observed at the localised emission band demonstrate preservation of the spin population in microsecond timescales **(f)**.

The indirect interband states and localized states are characterized by long exciton lifetimes. The recombination events require many-body processes due to the principle of preservation of momentum (e.g., recombination of an exciton with the emission of a phonon). Our measurements of PL decay times (**Fig. 3**) demonstrate microsecond lifetimes for both free indirect excitons and donor-bound excitons. More specifically, the description of the PL transients with a biexponential decay profiles yields lifetimes $\tau_1 = 0.10\ \mu s$ and $\tau_2 = 0.74\ \mu s$ for free exciton band and $\tau_1 = 0.21\ \mu s$ and $\tau_2 = 1.36\ \mu s$ for defect-bound exciton band. As expected, spatial localization effects of donor-bound excitons lead to an increase of the excitonic lifetime. The multicomponent characteristics of the PL decay time profiles arise most likely due to contribution of different excitonic complexes to the PL emission bands.

We also analyzed the spin properties of the charge carriers partaking in the optical processes by inspecting the magneto-optical effects in ZnPSe₃. Firstly, we observe the magnetic field evolution of the PL spectra resolved by circular polarization and excited resonantly in $\Gamma$ transition with circularly polarized light. The application of magnetic field creates a circular polarization degree of emitted photons for both PL bands, while no polarization degree is observable when exciting the sample with photons of opposite helicity (**Fig. 4(a,b)**). We can interpret this finding in a simple picture of magnetic-field induced splitting of individual valence and conduction bands that we describe with a splitting $g_{e,h}^{\Gamma(K)} \mu_B B$, where $g_{e,h}^{\Gamma(K)}$ is a g-factor for electron (e) or hole (h) in $\Gamma$ or K valley, $\mu_B$ is a Bohr magneton and B is the value of magnetic field. The relevant states in a presence of magnetic field, together with associated optical transitions, are illustrated in **Fig. 4(c)**. The values of the g-factors for individual bands are determined by the spin, orbital and/or valley effects[25,26]. No polarization degree observable for direct Γ excitation indicates that the oscillator strength for both σ⁺ and σ⁻ transition is independent on a magnetic field and the split valence states remain equally populated. In order to explain the emergent polarization degree associated with interband Γ-K recombination, we need to consider many-body excitonic states. For the simplest case of a neutral exciton, the initial state of the recombination process splits into σ⁺ and σ⁻ active components with a differential exciton g-factor $g_X^{\Gamma \to K} = g_e^{\Gamma} - g_h^K$. The final state of the recombination process is given by a magnetic-field-independent vacuum level. The magnetic field splitting of the excitonic states will induce their thermal repopulation manifesting as polarization degree of emitted photons. To investigate these effects quantitatively, we measured the polarization degree ($I_{\sigma^-}/I_{\sigma^+}$) for the interband exciton recombination as a function of magnetic field and temperature and we found that we may describe the polarization degree by a Boltzmann distribution function:

$$I_{\sigma^-}/I_{\sigma^+}(B,T) = A e^{\frac{g_X^{\Gamma \to K} \mu_B B}{k(T+T_{eff})}} + (1-A)$$

where $g_X^{\Gamma \to K}$ is the excitonic g-factor, $\mu_B$ is a Bohr magneton, $k$ is a Boltzmann constant, $T_{eff}$ is the effective temperature indicative of deviations from thermal equilibrium due to inter-excitonic interactions, $(1-A)$ accounts for remnant polarization degree ($I_{\sigma^-}/I_{\sigma^+}(B \to \infty, T)$) due to spin, orbital and/or valley mixing effects, $B$ and $T$ denote the value of magnetic field and sample temperature. Notably, we can describe the experimental data $I_{\sigma^-}/I_{\sigma^+}(B, 1.6\text{ K})$ and $I_{\sigma^-}/I_{\sigma^+}(9\text{ T}, T)$ with the same set of parameters (**Fig 4(d,e)**) and we found the best fit to yield $g_X^{\Gamma \to K} = 1.1 \pm 0.2$ and $T_{eff} = 2.6 \pm 1.0 K$. The agreement of our data with Boltzmann distribution and the low effective temperature indicates that the excitons in ZnPSe₃ in the investigates excitation conditions behave like weakly interacting semi-classical gas of particles[27-29]. Furthermore, polarization-resolved PL decay transients measured at 9 T for the donor-bound excitons demonstrate that the polarization degree is preserved throughout the microsecond lifetimes.

In conclusion, we have identified ZnPSe₃ as a van der Waals material with unique optical properties. We have demonstrated that ZnPSe₃ acts an indirect band-gap system that displays robust PL characterized by microsecond lifetimes of emitting states. Through a combination of PL and PLE spectroscopy we could selectively raise excitations within Γ and K valleys and probe the recombination processes of indirect interband transition and donor-bound excitons. The application of magnetic field allows repopulation of spin-split excitonic states opening avenues for spin manipulation in opto-electronic devices. Overall, ZnPSe₃ enriches the family of optically active 2D with potential to study valley-selective optical process and excitonic phase diagrams[30,31] in technologically relevant timescales.

**Acknowledgements**

This project was supported by the Ministry of Education (Singapore) through the Research Centre of Excellence program (grant EDUN C-33-18-279-V12, I-FIM). This material is based upon work supported by the Air Force Office of Scientific Research and the Office of Naval Research Global under award number FA8655-21-1-7026.

# References

[1]   M. Cardona and F. H. Pollak, Physical Review **142**, 530 (1966).
[2]   J. R. Schaibley, H. Yu, G. Clark, P. Rivera, J. S. Ross, K. L. Seyler, W. Yao, and X. Xu, Nature Reviews Materials **1**, 16055 (2016).
[3]   A. A. Kaplyanskii, N. S. Sokolov, B. V. Novikov, and S. V. Gastev, Solid State Communications **20**, 27 (1976).
[4]   K. S. Novoselov, D. Jiang, F. Schedin, T. J. Booth, V. V. Khotkevich, S. V. Morozov, and A. K. Geim, Proceedings of the National Academy of Sciences of the United States of America **102**, 10451 (2005).
[5]   A. H. Castro Neto, F. Guinea, N. M. R. Peres, K. S. Novoselov, and A. K. Geim, Reviews of Modern Physics **81**, 109 (2009).
[6]   F. Withers *et al.*, Nature Materials **14**, 301 (2015).
[7]   F. Withers *et al.*, Nano Letters **15**, 8223 (2015).
[8]   A. M. Jones *et al.*, Nature Nanotechnology **8**, 634 (2013).
[9]   G. Wang, L. Bouet, D. Lagarde, M. Vidal, A. Balocchi, T. Amand, X. Marie, and B. Urbaszek, Physical Review B **90**, 075413 (2014).
[10]  K. F. Mak, K. He, J. Shan, and T. F. Heinz, Nature Nanotechnology **7**, 494 (2012).
[11]  K. F. Mak, C. Lee, J. Hone, J. Shan, and T. F. Heinz, Physical Review Letters **105**, 136805 (2010).
[12]  Q. H. Wang, K. Kalantar-Zadeh, A. Kis, J. N. Coleman, and M. S. Strano, Nature Nanotechnology **7**, 699 (2012).
[13]  W. Zhao, Z. Ghorannevis, L. Chu, M. Toh, C. Kloc, P.-H. Tan, and G. Eda, ACS Nano **7**, 791 (2013).
[14]  A. Arora, M. Koperski, K. Nogajewski, J. Marcus, C. Faugeras, and M. Potemski, Nanoscale **7**, 10421 (2015).
[15]  A. Arora, K. Nogajewski, M. Molas, M. Koperski, and M. Potemski, Nanoscale **7**, 20769 (2015).
[16]  A. O. Slobodeniuk *et al.*, 2D Materials **6**, 025026 (2019).
[17]  C. Robert *et al.*, Physical Review B **93**, 205423 (2016).
[18]  T. Jakubczyk, V. Delmonte, M. Koperski, K. Nogajewski, C. Faugeras, W. Langbein, M. Potemski, and J. Kasprzak, Nano Letters **16**, 5333 (2016).
[19]  H. Xiang, B. Xu, Y. Xia, J. Yin, and Z. Liu, RSC Advances **6**, 89901 (2016).
[20]  W. S. Yun and J. D. Lee, The Journal of Physical Chemistry C **122**, 27917 (2018).
[21]  M. Sharma, A. Kumar, and P. K. Ahluwalia, AIP Conference Proceedings **1942**, 120019 (2018).
[22]  S. Rudin, T. L. Reinecke, and B. Segall, Physical Review B **42**, 11218 (1990).
[23]  K. P. O'Donnell and X. Chen, Applied Physics Letters **58**, 2924 (1991).
[24]  A. Hashemi, H.-P. Komsa, M. Puska, and A. V. Krasheninnikov, The Journal of Physical Chemistry C **121**, 27207 (2017).
[25]  M. Koperski, M. R. Molas, A. Arora, K. Nogajewski, A. O. Slobodeniuk, C. Faugeras, and M. Potemski, Nanophotonics **6**, 1289 (2017).
[26]  M. Koperski, M. R. Molas, A. Arora, K. Nogajewski, M. Bartos, J. Wyzula, D. Vaclavkova, P. Kossacki, and M. Potemski, 2D Materials **6**, 015001 (2018).
[27]  E. Hanamura and H. Haug, Physics Reports **33**, 209 (1977).
[28]  L. L. Chase, N. Peyghambarian, G. Grynberg, and A. Mysyrowicz, Physical Review Letters **42**, 1231 (1979).
[29]  D. Hulin, A. Mysyrowicz, and C. B. à la Guillaume, Physical Review Letters **45**, 1970 (1980).
[30]  P. L. Gourley and J. P. Wolfe, Physical Review Letters **40**, 526 (1978).
[31]  M. A. Tamor and J. P. Wolfe, Physical Review Letters **44**, 1703 (1980).

**Supplementary Information** for *"ZnPSe₃ as ultrabright indirect bandgap system with microsecond excitonic lifetimes."*

M. Grzeszczyk[1,2], K. S. Novoselov[1,2], M. Koperski[1,2,*]

[1]Department of Materials Science and Engineering, National University of Singapore, 117575, Singapore

[2]Institute for Functional Intelligent Materials, National University of Singapore, 117544, Singapore

E-mail: * msemaci@nus.edu.sg

1. **Sample preparation and experimental methods.**

Bulk crystals of ZnPSe₃ were grown via chemical vapor transport technique (by 2D Semiconductors). We cleaved the crystals along the plane perpendicular to 0001 direction of the crystal lattice and remove the top layers via mechanical exfoliation to obtain a clean crystal surface. Samples prepared via such method displayed stable and homogenous optical response in cryogenic and ambient conditions. No discernable degradation was observed within a month from the first experiments on cleaved crystals.

The optical characterization of ZnPSe₃ crystals was performed in backscattering microspectroscopic geometry. The sample was cooled down by a dry cryogenic system and the temperature was controlled by heaters at the sample and variable temperature insert (VTI) stages. The sample was positioned under an 50x objective by a stack of *x-y-z* piezo-positioners. The laser light was delivered into the low temperature insert through a monomode optical fiber for photoluminescence experiments and through a multimode 50 μm fiber for photoluminescence excitation experiments (PLE). The sample was excited by 405 nm single-frequency laser for monochromatic excitation and by supercontinuum source for PLE. In the latter case, the power was stabilized by a liquid crystal acting in a feedback loop with a photodiode probing the intensity of laser light after the output of the excitation fiber. The signal from the sample was collected through a multimode fiber, dispersed by 0.75 m spectrometer with a 150 g/mm grating and detected by a CCD camera. A set of filters, polarizers and waveplates was used to perform polarization-resolved experiments. The magnetic field was applied to the sample in the direction perpendicular to the layers of ZnPSe₃ material by superconducting coils. The photoluminescence decay times were measured by an avalanche photodiode under the excitation with picosecond 532 nm laser with kilohertz repetition rates.

2. **Identification of the donor-bound PL band.**

The low energy 2.00 eV emission band exists beyond the model of optical transition presented in the main text in **Fig. 1(e)**. In order to identify the origin of this band, we firstly notice that the PLE spectrum for the lower energy band is qualitatively similar to the recombination of the indirect interband states, which indicates that the excitation path also originates from photocreation of $X_\Gamma$ and $X_K$ states (see **Fig. 1(a,b)** in the main text). However, if we inspect the $X_\Gamma$ resonance, we notice that it is shifted towards lower energy by 25 meV and broadened by 36 % as compared to the $X_\Gamma$ resonance observed for the interband free exciton state. Based on these observations we conclude that the lower energy band corresponds to radiative recombination of donor bound excitons. Introduction of the lattice defect into the crystal structure can locally reduce the bandgap, which leads to the redshift of absorption resonances. Also, an inhomogeneous broadening is expected to occur as the excitons localized by the defects may initially be created at a distance away from the defect, hence the photoexcitated carriers statistically probe a variation of the bandgap from the minimum at the defect site to the maximum at the pristine lattice site.

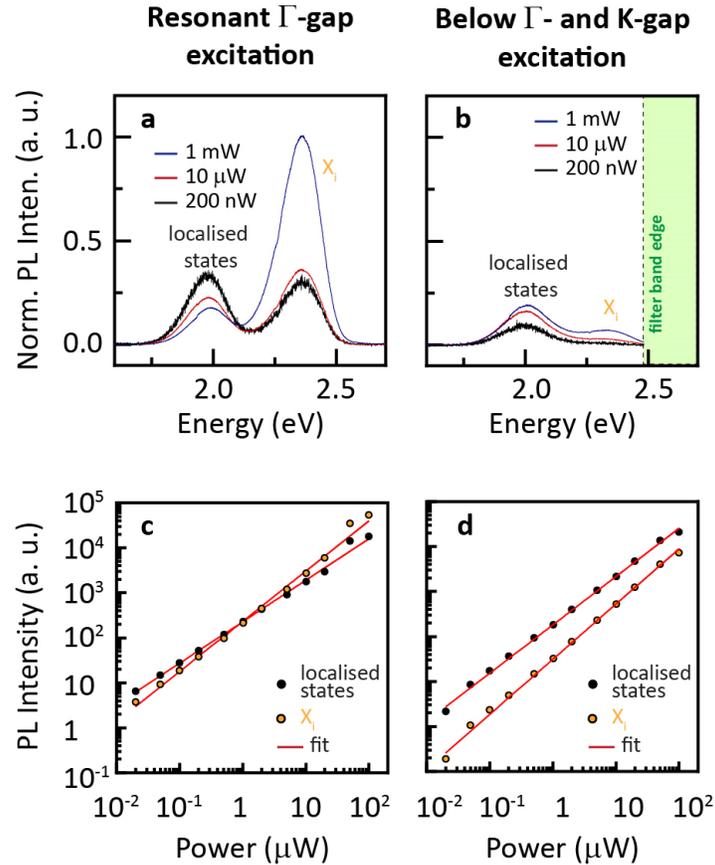

**Figure S1.** Laser excitation power dependence of the ZnPSe$_3$ PL spectra. The PL spectra are normalised by the excitation power and inspected in resonant excitation regime with the excitation energy matching the direct $\Gamma$ transition **(a)** as seen in the PLE spectrum and in non-resonant, sub-direct-gap excitation regime **(b)**. The detailed dependence of the PL intensity determined as the area under the PL bands across 4 orders of magnitude of excitation power is presented for resonant **(c)** and non-resonant excitations **(d)**.

We further corroborate this interpretation by inspecting the power dependence of the PL intensity comparatively in the resonant $\Gamma$-gap excitation regime and below direct gap excitation regime. The qualitative difference between the free interband excitons and defect-bound excitons manifests as a saturation of the latter band. Such finding is common in semiconductors, as the midgap defect states can trap a limited number of excitons when compared to the density of states of free interband excitons. This effect is clearly observable in **Fig. S1(a)**, where the PL spectra obtained under resonant $\Gamma$-gap excitation are normalized by the excitation power. In the low power regime, both bands of defect-bound excitons and free excitons are comparable in intensity. With the increase of power, the population of the defect-bound states saturates, while the intensity of the free exciton transitions grows superlinearly. In the regime of excitation below direct gaps (**Fig. S1(b)**), the free excitons band becomes significantly quenched, while the defect-bound states remain weakly excited through an alternative excitation process that is discernable in PLE spectra for the defect band as a broader resonance at 2.49 eV as illustrated in **Fig 1(b)** in the main text. This weaker excitation channel may be related to transitions from lower energy defect states directly into the conduction band. The detailed evolution of the PL intensity with the excitation power, highlighting the qualitative differences between resonant and under-gap excitation regimes, is presented in **Fig. S1(c,d)**.

### 3. Potential for photovoltaic applications.

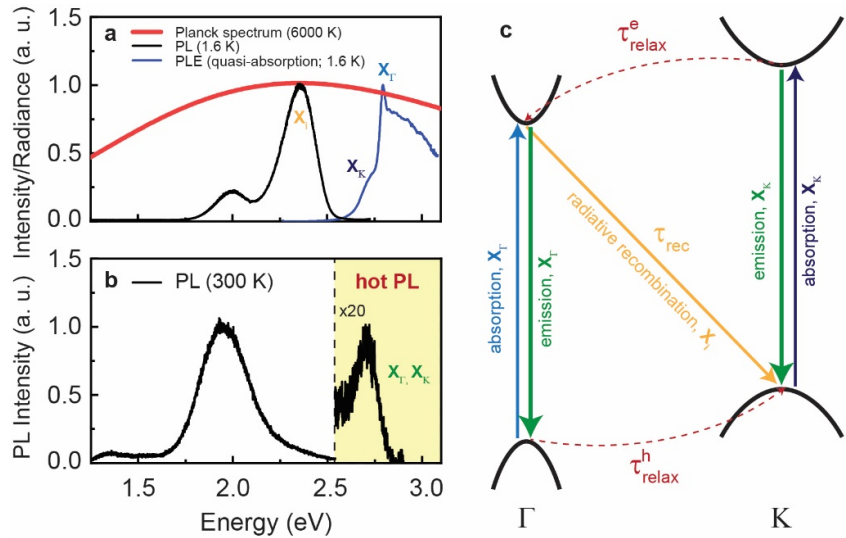

**Figure S2. (a)** The low temperature (1.6 K) PL and PLE spectra are compared with Planck spectrum at 6000 K characterising a spectral radiance of a black body that we treat as an approximation of a solar spectrum. **(b)** The emission spectra of ZnPSe$_3$ at room temperature exhibit signature of direct band gap recombination accompanying the ground state emission from the indirect excitonic state. The emergence of direct emission from and indirect band gap material is indicative of a competition between the radiative and relaxation processes of charge carriers in non-equilibrium conditions, as illustrated in **(c)** an extended schematic of optical processes in ZnPSe$_3$.

The optical activity of ZnPSe$_3$ occurs within the spectral region corresponding to the maximum radiance in the Planck spectrum for the temperature of the surface of the sun at about 6000 K as demonstrated in **Fig. S2(a)**. The energy of the interband transitions, in combination with microsecond excitonic lifetimes, makes ZnPSe$_3$ a reasonable candidate for photovoltaic applications. Long carrier lifetimes facilitate charge separation necessary for energy storage through the medium of photo-excited charge carriers. The single crystal character of ZnPSe$_3$ ensures homogenous optical response. The emergent defect centres predominantly trap excitons further increasing the recombination times rather than opening fast non-radiative channels. The favorable optical properties should open up pathways toward efficient solar cells among other energy storage optoelectronic devices.

### 4. Out-of-equilibrium excitonic states.

The optical response of ZnPSe$_3$ is dominated by ground state excitonic complexes at thermal equilibrium characterized by temperature close to the lattice temperature, as demonstrated in the main text. However, at room temperature the direct gap ($\Gamma$ and/or K) recombination is observable in PL spectra as presented in **Fig. S2(b)**. The manifestation of hot PL is indicative of the competition between direct recombination process and electron/hole relaxation process towards creation of the excitonic ground state (see **Fig. S2(c)** for a schematic illustration of these processes). Our data demonstrate that the relaxation process is dominant (i.e., occurs in shorter timescale), nevertheless the recombination of hot, excited excitonic states is still observable at room temperature.

## 5. Exciton binding energy.

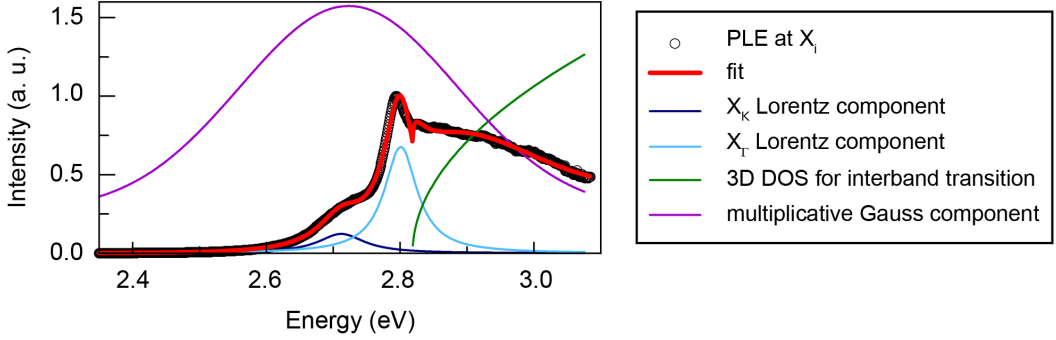

**Figure S3.** The low temperature PLE spectrum monitored at the energy of indirect free exciton transition ($X_i$) is deconvoluted into two excitonic resonances ($X_K$ and $X_\Gamma$) and an interband single particle absorption edge as additive components. The entire spectrum is multiplied by a Gaussian broadband function to account for the wavelength dependent focus of the objective leading to modulation of the collected optical signal.

Our analysis of the quasi-absorption PLE spectra allows us to provide an estimation of the $\Gamma$ and K exciton binding energies. As presented in **Fig. S3**, our fitting has three major components: 1) $\Gamma$ exciton resonance, 2) K exciton resonance and 3) three dimensional density of states characterizing the single particle interband absorption process. We treat these three terms additively, while introducing a broadband multiplicative Gaussian component to account for chromaticity of the focal distance of the objective in PLE experiments. Overall, the fitting function has the following formulation:

$$I(E) = \begin{cases} \left( I_{0,K} \dfrac{\Delta E_K}{(E-E_K)^2 + \left(\Delta E_K/2\right)^2} + I_{0,\Gamma} \dfrac{\Delta E_\Gamma}{(E-E_\Gamma)^2 + \left(\Delta E_\Gamma/2\right)^2} + I_{0,DOS}\sqrt{E-E_{gap}} \right) * \left( I_{0,Gauss} * e^{-\frac{(E-E_0)^2}{2\sigma^2}} + B \right), \text{for } E \geq E_{gap} \\ \left( I_{0,K} \dfrac{\Delta E_K}{(E-E_K)^2 + \left(\Delta E_K/2\right)^2} + I_{0,\Gamma} \dfrac{\Delta E_\Gamma}{(E-E_\Gamma)^2 + \left(\Delta E_\Gamma/2\right)^2} \right) * \left( I_{0,Gauss} * e^{-\frac{(E-E_0)^2}{2\sigma^2}} + B \right), \text{for } E < E_{gap} \end{cases}$$

From mean square error fitting, we find the energy values to be $E_K = 2.711 \pm 0.005$ eV, $E_\Gamma = 2.802 \pm 0.003$ eV and $E_{gap} = 2.821 \pm 0.005$ eV. As we cannot discern two separate absorption band edges at K and $\Gamma$ point, we will assume the $E_{gap}$ energy corresponds to lower energy single particle $\Gamma$ band gap (according to density functional theory calculations), while K point gap is expected to be slightly larger.

In such a case, we can provide estimates for the exciton binding energy to be $E^K_{binding} = 110 \pm 10$ meV and $E^\Gamma_{binding} = 19 \pm 8$ meV. It is important to note, this in this analysis the value of binding energy for the K exciton is underestimates by the difference in the single particle band gap between K and $\Gamma$ points.

The linewidths for K and $\Gamma$ resonances, which are used to calculate the relaxation times of electrons and holes in the main texts, were found to be $\Delta E_K = 91 \pm 4$ meV and $\Delta E_\Gamma = 64 \pm 3$ meV.